\documentclass[
    ,final            
  ]
  {aipproc}

\layoutstyle{6x9}
\def\gsim{\mathrel{\raise0.35ex\hbox{$\scriptstyle >$}\kern-0.6em
\lower0.40ex\hbox{{$\scriptstyle \sim$}}}}
\def\lsim{\mathrel{\raise0.35ex\hbox{$\scriptstyle <$}\kern-0.6em
\lower0.40ex\hbox{{$\scriptstyle \sim$}}}}

\begin{document}

\title{Starburst or AGN Dominance in Submillimetre-Luminous Candidate AGN?}


\author{Kristen Coppin}{
  address={ICC, Durham University, UK}
}

\author{Alexandra Pope}{
  address={\textit{Spitzer Fellow}, National Optical Astronomy
    Observatory, Tuscon, AZ, USA}
}

\author{Kar\'{i}n Men\'{e}ndez-Delmestre}{
  address={NSF Astronomy and Astrophysics Postdoctoral Fellow, the
    Observatories of the Carnegie Institution for Science, Pasadena,
    CA, USA}
}

\author{David M. Alexander}{
  address={Department of Physics, Durham University, UK}
}

\author{James Dunlop}{
  address={Scottish Universities Physics Alliance, Institute
    for Astronomy, University of Edinburgh, Royal Observatory, UK}
}

\begin{abstract}
It is widely believed that ultraluminous infrared (IR) galaxies and active galactic
nuclei (AGN) activity are triggered by galaxy interactions and
merging, with the peak of activity occurring at $z\sim2$, where
submillimetre galaxies are thousands of times more numerous than local
ULIRGs.  In this evolutionary picture, submillimetre galaxies (SMGs)
would host an AGN, which would eventually grow a black hole (BH) strong
enough to blow off all of the gas and dust leaving an optically
luminous QSO. To probe this evolutionary sequence we have
focussed on the 'missing link' sources, which demonstrate both strong
starburst (SB) and AGN signatures, in order to determine if the SB
is the main power source even in SMGs when we have evidence that an
AGN is present from their IRAC colours.  The best way to determine if a dominant AGN is
present is to look for their signatures in the mid-infrared with the \textit{Spitzer} IRS, since
often even deep X-ray observations miss identifying the presence of
AGN in heavily dust-obscured SMGs. We present the results of our audit
of the energy balance between star-formation 
and AGN within this special sub-population of SMGs -- where the BH 
has grown appreciably to begin heating the dust emission.
\end{abstract}

\maketitle


\section{Main Results}
\textit{Spitzer} spectroscopy has revealed that $\simeq80$\% of SMGs
are SB-dominated in the mid-infrared (\cite{Pope08};
\cite{KMD09}).  We have focussed on the remaining $\simeq20$\% that
show signs of harboring powerful AGN and use \textit{Spitzer}-IRS
spectroscopy to study a sample of 8 SMGs $>200\,\mu$Jy at 24$\,\mu$m from the
SCUBA Half Degree Extragalactic Survey (SHADES; \cite{Coppin06})
selected on the basis of an IRAC color-selection
($S_{8\mu\mathrm{m}}/S_{4.5\mu\mathrm{m}}>2$; i.e.~likely power-law
mid-infrared SEDs; see Fig.~1).  The full analysis will be presented in Coppin et al.\
(ApJ, submitted), and our main results are as follows:
\begin{itemize}
\item There are signs of SF from PAH features in \textit{all} of our 
SMGs, from which we derive redshifts between 2.5--3.4,
demonstrating the power of the mid-IR to determine redshifts
when the optical counterparts are too faint to study with current
facilities.  
\item  The IRS spectra show signs of \textit{both} SB and AGN
  activity in our $S_{8\mu\mathrm{m}}/S_{4.5\mu\mathrm{m}}>2$ SMG sample, with a
continuous distribution of AGN fractions in the mid-IR.  Overall, SMGs
selected in this way tend to have more dominant AGN-components in the
mid-IR than typical SMGs, with a median AGN-fraction of
58\%. Although, extrapolation to the far-IR reveals that the AGN
is bolometrically unimportant in the majority of SMGs, indicating that the
level of AGN contamination in the overall SMG population is probably
$\lsim5$\%.  For comparison, typical SMGs have $<30$\% AGN
contribution in the mid-infrared, while they are from a very similar
SB $L_\mathrm{IR}$ class to our sample of
$\simeq5\times10^{12}\,\mathrm{L}_\odot$. 
\item When literature sources are taken into account, a
  colour-selection of $S_{8}/S_{4.5}>1.65$ is a better description
  overall for defining the boundary between SB and AGN-dominated SMGs,
  with a small amount of scatter across this division (see Fig.~1).  
\item Our results are thus consistent with the evolutionary scenario \cite{Sanders88}, with all SMGs undergoing a `transitional' AGN-dominated phase with a duty cycle of $\simeq20$\%. Our sample of AGN-dominated SMGs could be at a slightly later stage of evolution than SF-dominated SMG systems, with the SF still occurring but where the AGN has now begun to heat the dust appreciably in the SMG as the BH undergoes a period of rapid growth.
\end{itemize}





\begin{minipage}{2.8in}
\hspace{-0.2in}
\includegraphics[height=.28\textheight]{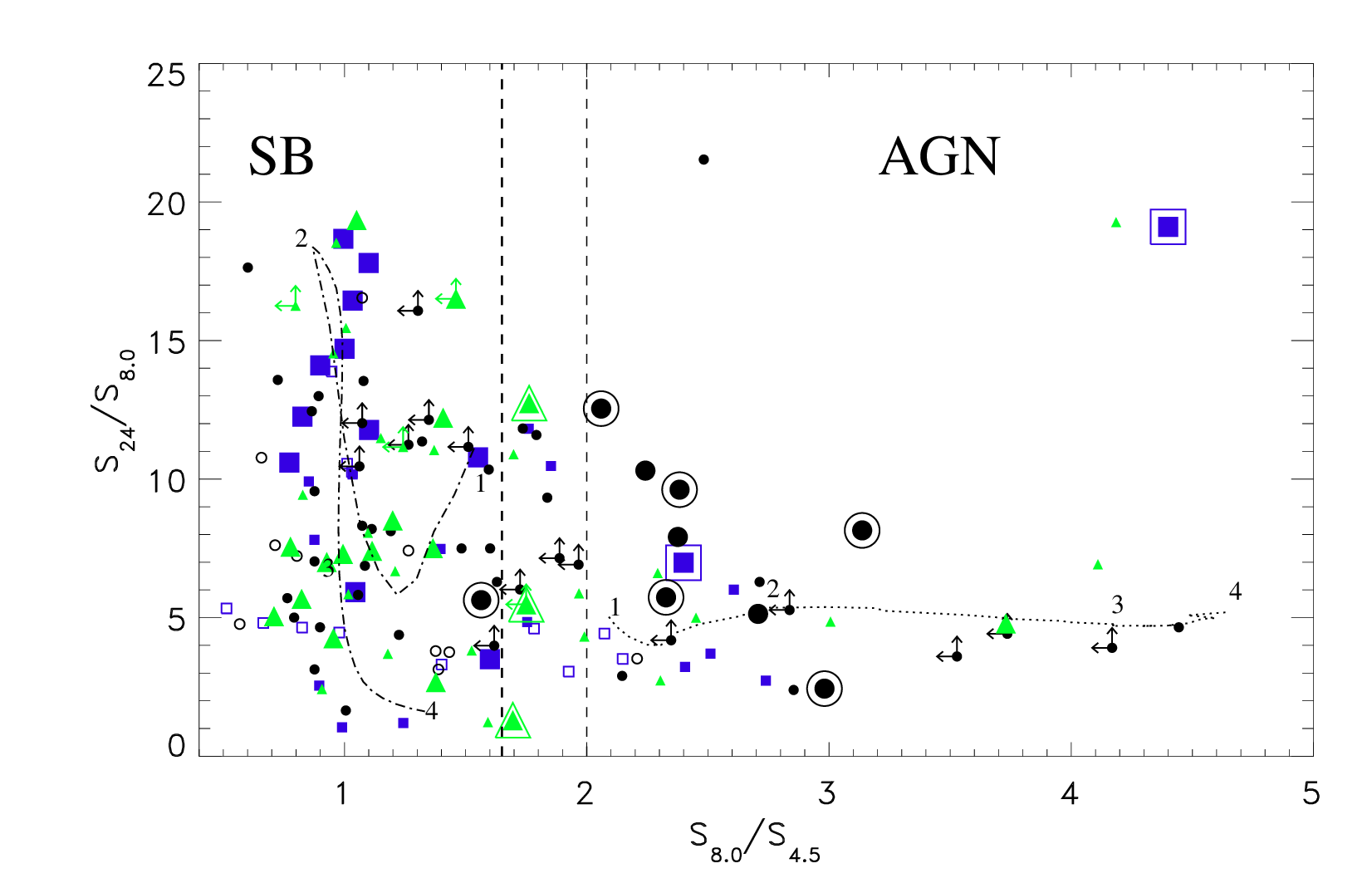}
\end{minipage}
\hspace{0.6in}
\begin{minipage}{2.in}
{\small{Figure 1:~~{\it Spitzer} colour-colour diagram as an AGN
    diagnostic for SMGs.  Circles are SHADES SMGs, and other samples
    are plotted as squares \cite{Pope08} and triangles \cite{KMD09}.  Mrk231 (an AGN; dotted line)
  and M82 (a SB; dot-dashed line) are plotted as a function of
  redshift for comparison.   Sources with a $>50\%$ contribution from continuum (AGN)
  emission to their mid-IR emission are \textbf{outlined} and tend to
  lie far from the M82 track.}}
\end{minipage}

\begin{theacknowledgments}
We would like to thank the conference organizers, STFC, NSF, NASA and
the Royal Society for financial support, and also our collaborators from SHADES who
contributed to this work.  This work is based in part on observations made with the \textit{Spitzer Space Telescope}, which is operated by the Jet Propulsion Laboratory, California Institute of Technology under a contract with NASA. 
\end{theacknowledgments}



\bibliographystyle{aipproc}   

\begin{thebibliography}{}

\bibitem[Coppin et al.(2006)]{Coppin06}Coppin K., et al., 2006, \emph{MNRAS}, 372, 1621

\bibitem[Men\'{e}ndez-Delmestre et al.(2009)]{KMD09} Men\'{e}ndez-Delmestre K., et al., 2009, \emph{ApJ}, 699, 667

\bibitem[Pope et al. (2008)]{Pope08} Pope A., et al., 2008, \emph{ApJ}, 675, 1171

\bibitem[Sanders et al.(1988)]{Sanders88}Sanders D.B., Soifer B.T.,
  Elias J.H., Neugebauer G., Matthews K., 1988, \emph{ApJ}, 328, L35

\end{thebibliography}


\end{document}